\documentclass{article}
\usepackage{icmc2025_paper_template}
\usepackage{times}
\usepackage{ifpdf}
\usepackage{soul}
\usepackage[english]{babel}

\newcommand{\fo}{$f_0$}

\def\papertitle{pyAMPACT: A Score-Audio Alignment Toolkit for Performance Data Estimation and Multi-modal Processing}
\def\firstauthor{Johanna Devaney}
\def\secondauthor{Daniel McKemie}
\def\thirdauthor{Alexander Morgan}

\newif\ifpdf
\ifx\pdfoutput\relax
\else
   \ifcase\pdfoutput
      \pdffalse
   \else
      \pdftrue
  \fi
\fi

\ifpdf % compiling with pdflatex
  \usepackage[pdftex,
    pdftitle={\papertitle},
    pdfauthor={\firstauthor, \secondauthor, \thirdauthor},
    bookmarksnumbered, % use section numbers with bookmarks
    pdfstartview=XYZ % start with zoom=100% instead of full screen; 
   ]{hyperref}
  \usepackage[pdftex]{graphicx}
  \graphicspath{{./figures/}}
  \DeclareGraphicsExtensions{.pdf,.jpeg,.png}

  \usepackage[figure,table]{hypcap}

\else % compiling with latex
  \usepackage[dvips,
    bookmarksnumbered, % use section numbers with bookmarks
    pdfstartview=XYZ % start with zoom=100% instead of full screen
  ]{hyperref}  % hyperrefs are active in the pdf file after conversion

  \usepackage[dvips]{epsfig,graphicx}
  \graphicspath{{./figures/}}
  \DeclareGraphicsExtensions{.eps}

  \usepackage[figure,table]{hypcap}
\fi

\hypersetup{
    colorlinks,%
    citecolor=black,%
    filecolor=black,%
    linkcolor=black,%
    urlcolor=black
}

\title{\papertitle}

 \threeauthors
   {0.5in}
   {\firstauthor} {Brooklyn College, CUNY \\ %
     {\tt \href{mailto:jcdevaney@gmail.com}{jcdevaney@gmail.com}}}
   {\secondauthor} {Brooklyn College, CUNY \\ %
     {\tt \href{mailto:daniel.mckemie@gmail.com}{daniel.mckemie@gmail.com}}}
   {\thirdauthor} { Independent \\ %
     {\tt \href{mailto:alexanderpmorgan@gmail.com}{alexanderpmorgan@gmail.com}}}

\begin{document}
\capstartfalse
\maketitle
\capstarttrue
\begin{abstract}
\texttt{pyAMPACT} (Python-based Automatic Music Performance Analysis and Comparison Toolkit) links symbolic and audio music representations to facilitate score-informed estimation of performance data from audio. It can read a range of symbolic formats and can output note-linked audio descriptors/performance data into MEI-formatted files. \texttt{pyAMPACT} uses score alignment to calculate time-freq-uency regions of importance for each note in the symbolic representation from which it estimates a range of parameters from the corresponding audio. These include frame-wise and note-level tuning-, dynamics-, and timbre-related performance descriptors, with timing-related information available from the score alignment. Beyond performance data estimation, \texttt{pyAMPACT} also facilitates multi-modal investigations through its infrastructure for linking symbolic representations and annotations to audio.
\end{abstract}

\section{Introduction}\label{sec:introduction}

\texttt{pyAMPACT} uses score-audio alignment to facilitate expressive performance modeling. Although this approach requires a score, it provides more robust estimates of onsets and offsets than automatic transcription methods.  Although there have recently been significant improvements in the accuracy of automatic transcription, largely due to advances in deep-learning architecture, accuracy on this task is still limited \cite{benetos_automatic_2018,bittner_lightweight_2022}. Symbolic representations also contain additional information that can also be difficult to accurately estimate solely from the audio signal, including meter and note spellings, which is useful for musicological analysis of performance data. Beyond expressive performance, \texttt{pyAMPACT}'s tools for linking symbolic and audio representations are useful for multi-modal processing of audio and score-based representations with annotations and other types of human-generated data. \texttt{pyAMPACT} is based on \texttt{AMPACT}, a MATLAB-based toolkit originally released in 2011 \cite{devaney11automatically,devaney12ampact}. It is not simply a Python-based re-implementation of \texttt{AMPACT}, but rather a complete redesign that connects to and extends the functionality of an array of other open-source music analysis tools, most notably librosa \cite{mcfee2015librosa} and music21 \cite{cuthbert2010music21}. It also moves away from the proprietary nature of MATLAB while building on the signal-processing algorithms integrated into \texttt{AMPACT} \cite{devaney09improving, devaney11characterizing, devaney17evaluation, devaney19encoding}. \texttt{pyAMPACT} is also able to read a range of annotation encoding formats, including Dezrann \cite{giraud18dezrann},  Humdrum's analytic spines \cite{huron02music}, and CRIM intervals \cite{freedman22musicologists}). 
It can also visualize both score and performance data in Verovio \cite{pugin14verovio}.

In this paper, we first contextualize \texttt{pyAMPACT} within related work on score-audio alignment, performance data estimation, and multi-modal processing (Section 2). We then present an overview of \texttt{pyAMPACT}’s workflow (Section 3) before detailing its symbolic importing (Section 4),  audio processing (Section 5), and symbolic exporting (Section 6) functionality, as well as its documentation and tutorials (Section 7). We conclude with a summary of \texttt{pyAMPACT} and a discussion of future directions (Section 8).

\section{Related Work}\label{sec:background}

Estimation of performance parameters from polyphonic audio requires accurate extraction of frequency and power information for each note's fundamental frequency and all of its partials. While deep-learning approaches in the polyphonic note-transcription have improved the state-of-the-art, there remains a performance ceiling for both note event detection and \fo{} accuracy \cite{benetos_automatic_2018,bittner_lightweight_2022}. An alternative, which has been explored since the 1990s \cite{scheirer1998using}, is to use score-audio alignment to inform the estimation of performance parameters. Standard algorithms estimate a single time point estimate for the start of each notated simultaneity \cite{orio_alignment_2001,raphael2004hybrid} while some algorithms estimate onset and offset for each note in a notated simultaneity \cite{niedermayer2010multi,devaney14estimating,carabias2015audio}. As with most other MIR tasks, deep-learning approaches have been applied to score-audio alignment \cite{simonetta2021audio, agrawal2021convolutional}, including work with the goal of improving multi-pitch estimation with weakly aligned scores \cite{krause2023soft}. 

Performance parameter estimation is valuable for expressive performance modeling, which traditionally has been done for the purposes of both analysis and generation \cite{kirke2012guide,cancino2018computational,lerch2020interdisciplinary}. Work on expressive performance is still largely focused on the piano \cite{shi2021computational, simonetta2022acoustics,zhang2022atepp}, although the voice and other instruments are increasingly being studied \cite{giraldo2020machine,wang_pipaset_2022,  dai2023singstyle111,tamer2023high}. Some recent work on performance parameter estimation through score-audio alignment has also examined the use case of detecting errors in performance, both to improve the score-audio alignment accuracy \cite{nakamura2017performance, li2022regularizedlian2023pqg} and to assess performance \cite{huang2019automatic,li2022regularized}. Score-audio alignment has also been leveraged for multi-modal processing \cite{foscarin2020asap} and producing large-scale multi-modal datasets themselves \cite{meseguer2020creating, simonetta2020asmd, weiss2023wagner}. This helps to address the data paucity issue facing multi-modal processing \cite{simonetta_multimodal_2019}.

\begin{figure*}
    \centering
    \includegraphics[width=\linewidth]{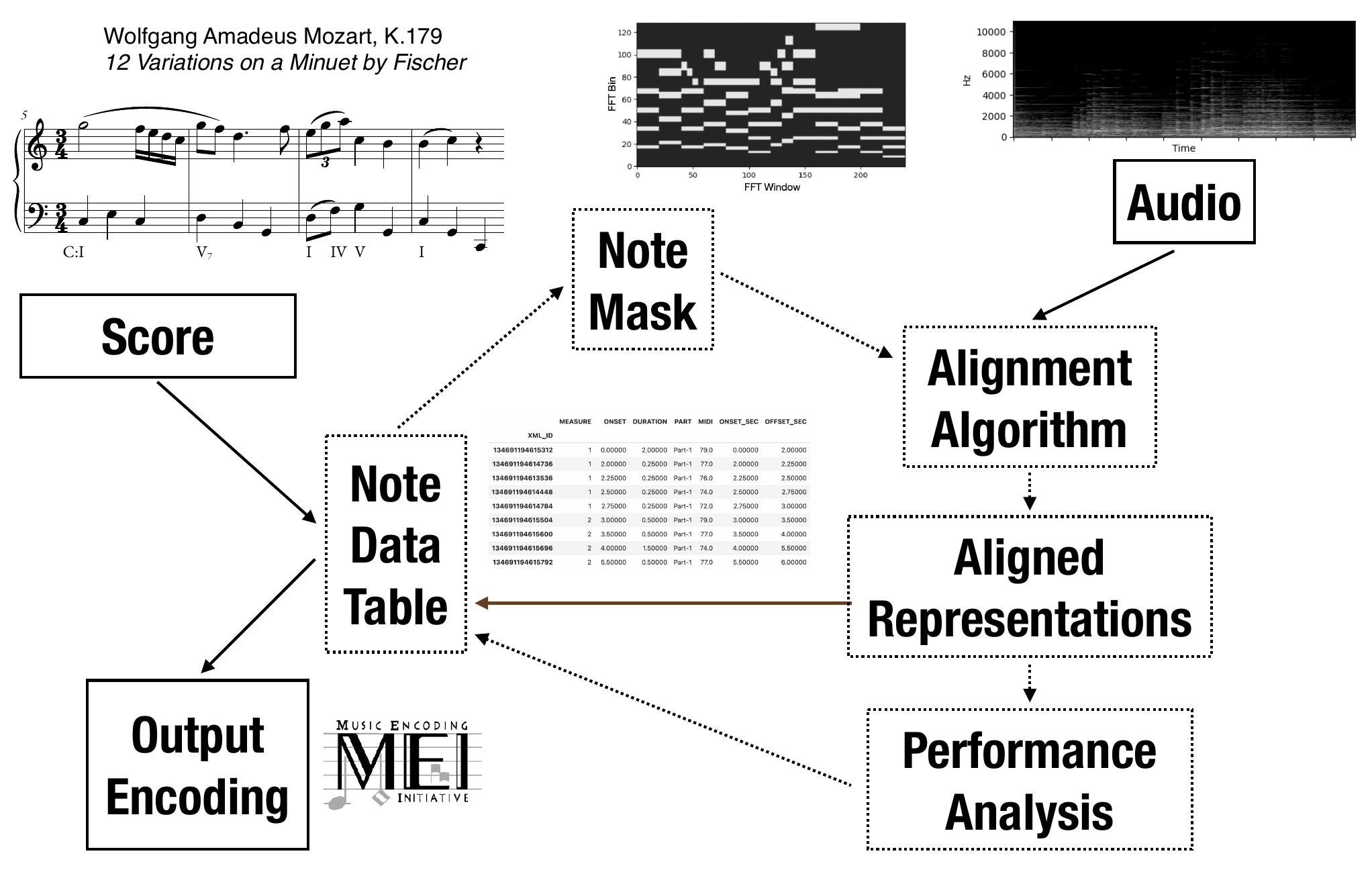}
    \caption{Overview of \texttt{pyAMPACT}'s workflow.}
    \label{fig:overview}
\end{figure*}

\section{Overview of \texttt{pyAMPACT}'s Workflow}\label{workflow}

Figure \ref{fig:overview} provides an overview of \texttt{pyAMPACT}'s workflow. Symbolic data ('Score') along with any time-aligned annotations are imported and stored in a series of \texttt{Pandas} DataFrames (including a 'Note Data Table'). The score-based events in the DataFrames are linked to the original symbolic representation. This allows for the analysis and export of the estimated performance data. The score data stored in the 'Note Data Table' is converted into a spectrogram-like mask ('Note Mask') to facilitate alignment with the imported audio. The 'Alignment Algorithm' calculates a mapping between each note in the symbolic representation and specific time-frequency regions in the audio spectrogram ('Aligned Representations'). The alignment is currently performed by the DTW-based algorithm used in \texttt{AMPACT}, we plan to make this extensible, so other alignment algorithms can be used. 
The timing information in the 'Note Data Table' is updated with the timing estimates from the alignment stored in the 'Aligned Representations'. The note-wise time-frequency regions estimated through the alignment process are used to estimate tuning-, dynamics-, and timbre-related performance parameters ('Performance Analysis'). The estimated frame- and note-level performance parameters are also stored in the 'Note Data Table', which facilitates the analysis of the performance data in relation to symbolic data and any linked annotations. The performance data in the 'Note Data Table' can either be analyzed within Python or exported into a standard encoding format (i.e., MEI) linked with the symbolic events in the original imported symbolic data ('Output Encoding') for analysis in other coding environments.

%This is facilitated by \texttt{pyAMPACT}'s tools for not only aligning note information in imported symbolic representations but also any note-linked analytic annotations. \texttt{pyAMPACT} such as \texttt{Dezrann} or \texttt{Humdrum} kern. It also facilitates exporting the linked symbolic and performance data into a new symbolic file, in either MEI or \texttt{Humdrum} kern format, (f) in Figure \ref{fig:overview}.}

\section{Importing Symbolic Data}\label{sec:symbolic}
\texttt{pyAMPACT} imports scores through an exposed music21 score object representation, allowing it to leverage music21 support all major symbolic notation file types (e.g.,  Humdrum \verb|**kern|, MEI, MIDI, and MusicXML), as well as less commonly used ones (e.g., ABC, TinyNotation, and Volpiano).  This is an expansion of file formats from the original \texttt{AMPACT}, which only supports MIDI files. 

Symbolic data is accessible through \texttt{pyAMPACT}'s Score class, which encodes the following information for each note in the symbolic representation: XML ID (for linking to imported score data), measure number, onset and duration in beats (measured from the beginning of the piece), part number, MIDI note number, and onset and offset times in seconds. The onset and offset times are originally populated with a placeholder value based on 60 bpm. These values are subsequently updated by the alignment algorithm with times from the aligned audio file.

To calculate accurate durations from the imported file, it is first converted to a temporary MIDI file and parsed by track. Each track's messages are processed, with types read for tempo, note-on, note-off, and end-of-track events. If tempo information is defined in the source file, it is used throughout; otherwise, a default of 60 BPM is assigned. Note-on and note-off messages with a velocity greater than 0 (to ignore rests) are captured and used to calculate the start and end times of each note, based on pulses per quarter note (PPQN)\footnote{Start time is calculated using \texttt{start\_time = current\_tempo / (1\_000\_000 * ppqn)}, where the tempo is converted to microseconds to determine the pulses. These calculations are cumulatively added to determine the \texttt{ONSET\_SEC} time of each note. Similarly, the note-off messages follow the same process to calculate the end time \texttt{(OFFSET\_SEC)}. Finally, the duration of each note is determined as \texttt{DURATION = OFFSET\_SEC - ONSET\_SEC}.}.

\begin{figure*}[h]
    \centering
    \includegraphics[width=\linewidth]{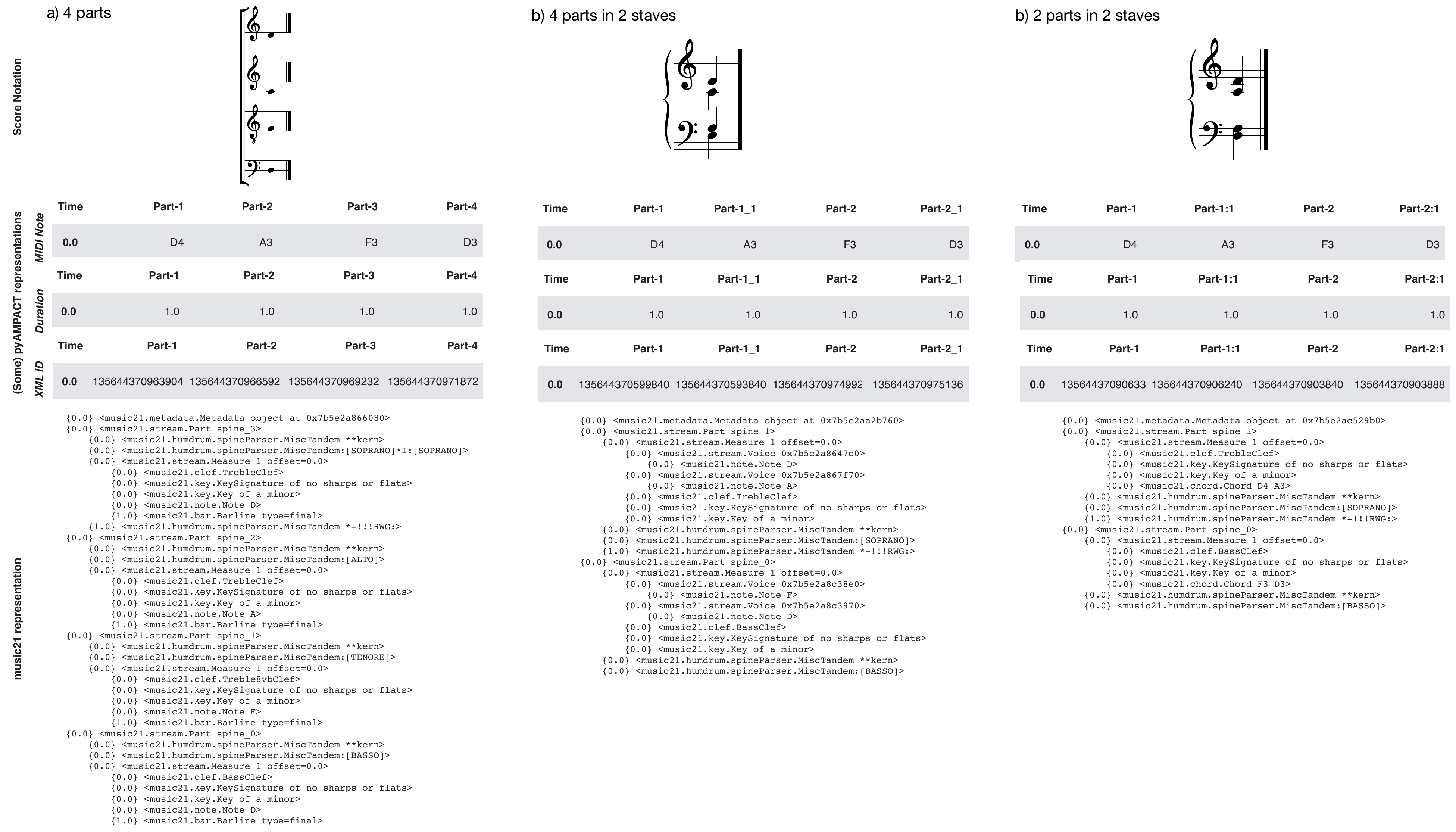}
    \caption{Musical score data (upper row) represented in DataFrames by \texttt{pyAMPACT} (middle row) and as trees by \texttt{music21}. Subplot (a) shows a chord voiced as 4 separate voices, each in their own staves. Subplot (b) shows the same chord voiced as 4 separate voices, grouped into two staves. Subplot (c) shows the same chord with the notes in a single voice in each of the two staves.}  
    \label{fig:reps}
\end{figure*}

When importing symbolic files, \texttt{pyAMPACT} converts \texttt{music21}’s stream-based representation of score data in Python lists to a tabular representation. Specifically as \texttt{Pandas} DataFrames which allows users to interact with the core \texttt{pyAMPACT} symbolic representation in a standardized way. In these DataFrames, the \texttt{music21} symbolic note data is simplified to events and timings: score components are represented as columns and score time as rows, with score time measured in \texttt{music21} offsets (where an eighth note is 0.5, a quarter note 1, a half note 2, etc.)\footnote{This basic format, with score parts on the x-axis and time on the y-axis, is similar to Humdrum kern. However, unlike Humdrum kern where notes, measures, and time signatures are all in one place, each cell in a \texttt{pyAMPACT} DataFrame typically corresponds to a single type of information.} Separate \verb|nmat| DataFrames are generated for each audio file aligned to a symbolic representation and populated with estimated performance data. Figure \ref{fig:reps} shows the compactness of the DataFrame-based representation used in \texttt{pyAMPACT} versus the tree-based stream representation used in \texttt{music21}, particularly across different voicings of the musical material.

The Score object has several methods which output different representations of the data in the symbolic notation file, including a piano roll, a note data table (\verb|nmat|) and a spectrogram-like mask (verb|mask|). The note data table, or note matrix, (\verb|nmat|) DataFrame is based on the representation in the MIDI Toolbox  \cite{eerola04mir}. The \verb|mask| DataFrame, based on Dan Ellis’ \verb|alignmidiwav| code \cite{ ellis13align}, provides a spectrogram-like representation of the symbolic data to align with a spectrogram of an audio recording. In addition to facilitating the audio-score alignment necessary for estimating performance parameters within \texttt{pyAMPACT}, the alignment of the score and audio representations generated by \texttt{pyAMPACT} is useful for multi-modal processing more generally.

\begin{figure*}[h]
  \centering
  \includegraphics[width=\linewidth]{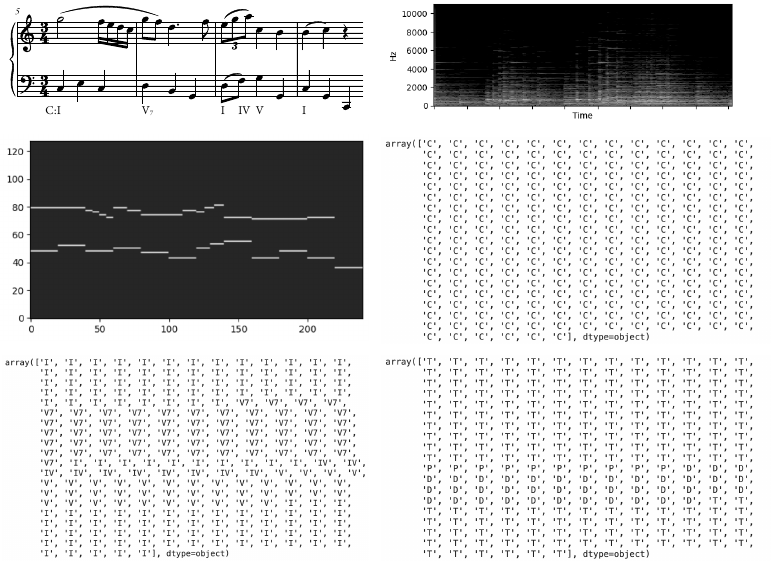}
  \caption{Annotations.} 
\label{fig:annotations}
\end{figure*}

%\subsection{Importing Annotations}\label{annotations}

\subsection{Multi-modal Processing}\label{sec:multimodal}

The representations shown in Figure \ref{fig:reps} are some of the possible time-aligned representations in \texttt{pyAMPACT}. The symbolic and spectrogram representations at the top of the figure are produced from the input score and audio representations. The lower part of the figure shows the piano roll (\verb|pianoroll|), Roman numeral, pop chord, and harmonic function representations, a spectrogram-like mask of the symbolic representation (as shown in Figure \ref{fig:overview}) is also available. This is facilitated by \texttt{pyAMPACT}'s support for importing aligned analytic annotations from a range of established formats (including Humdrum and Dezrann). 

The Humdrum ecosystem includes numerous established and highly capable analytical methods with results encoded in a variety of spine types. \texttt{pyAMPACT} can directly read in several Humdrum spine types ( \verb|**harm|, \verb|**chord|, and \verb|**function|) and offers a generalized solution for importing any user-defined spine type found in a Humdrum \verb|**kern| file. This allows users to take a working Humdrum analysis workflow, and import its output into pyAMPACT where it is possible to engage with audio analysis or other symbolic analysis available in pyAMPACT. Analytic annotations from .dez files are also supported. In addition to exploring connections between analyses from .dez files and corresponding audio observations from pyAMPACT, pyAMPACT users benefit from the ability to use Dezrann's feature-rich and web-based platform as a graphic interface for precisely annotating scores. \texttt{pyAMPACT}.

In addition to the analytic tools in music21, pyAMPACT can also integrate analytic annotations from the Citations: the Renaissance Imitation Mass (\texttt{CRIM}) suite of symbolic analysis. pyAMPACT's data structuring is sufficiently similar to CRIM Intervals that users can directly compare analysis results from a CRIM Intervals method with audio analysis results from pyAMPACT. In the event that an imported score does not have any Roman Numeral annotations, \texttt{pyAMPACT} runs the relevant music21 analysis methods to estimate the Roman Numerals.

%Roman numeral chord analysis stored in a \texttt{pyAMPACT} Score object can be accessed by calling \verb|romanNumerals()| on that object regardless of which supported format the data was imported from. This also means that annotations imported from different formats (such as Humdrum \verb|**harm| spines and \verb|.dez| Roman numeral analysis) are returned in the same time-aligned format. 

\section{Audio Processing}\label{sec:audio}

\begin{figure*}[h]
  \centering
  \includegraphics[width=\linewidth]{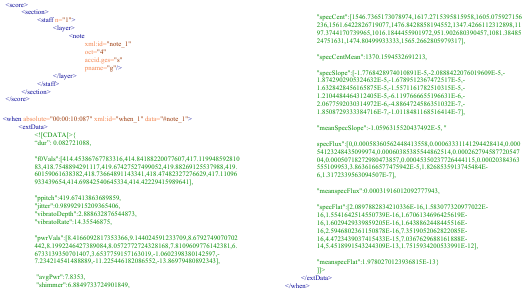}
  \caption{Example of performance data encoded in MEI using the \texttt{<extData>} tag. The performance data is linked to the corresponding note in the \texttt{<score>} tag through the \texttt{data} attribute of the \texttt{<when>} tag. The performance data is encoded as a JSON object in a \texttt{<![CDATA[>} tag.} 
\label{fig:extData}
\end{figure*}

% Is the HMM worth mentioning in Future Work?
The imported symbolic data is aligned to a corresponding audio file to estimate note onset and offset times, which replace the initial place-holder onset and offset times in the NMAT representation (described in Section \ref{sec:symbolic}). These onset and offset estimates identify time-frequency regions of importance, which facilitates performance parameter estimation in both monophonic and polyphonic audio. This score-guided approach to performance data estimation is implemented in the original \texttt{AMPACT} toolkit and described in \cite{devaney17evaluation}. The basic score alignment implemented in \texttt{pyAMPACT} is a standard dynamic time warping (DTW) approach \cite{orio2001alignment}, however, \texttt{pyAMPACT} is extensible so that any Python-based score-alignment can be used be used. Once the symbolic data has been aligned to the audio data, a number of frame-level descriptors are estimated per aligned note and also summarized into a set of note-level descriptors. An example of the performance data estimated for a single note is shown in Figure \ref{fig:extData}.

% Example of MEI score data encoding. In this example the \texttt{oct} (octave), \texttt{accid.ges} (accidental), and \texttt{pname} (pitch) properties were derived from the note estimates in Tony and the \texttt{xml:id} attribute allows linking with the performance data in Figure \ref{fig:extData}.
%Example of MEI score data encoding. In this example the \texttt{oct} (octave), \texttt{accid.ges} (accidental), and \texttt{pname} (pitch) properties were derived from the note estimates in Tony and the \texttt{xml:id} attribute allows linking with the performance data in Figure \ref{fig:extData}

\subsection{Frame-Level Descriptors}
\texttt{pyAMPACT} uses an instantaneous frequency representation, produced by librosa's reassigned\_spectrogram function, to estimate frame-level fundamental frequency (\fo{}) and power. A harmonic spectral representation is calculated from the fundamental frequency and power estimates. A set of frame-level descriptors (currently spectral centroid spectral slope, and spectral flux). 

\subsection{Note-Level Descriptors}
From the frame-wise estimates of \fo{}, power, spectral descriptors described above, \texttt{pyAMPACT} currently estimates five pitch-related note-level descriptors, two dynamics-related note-level descriptors, and four timbre-related note-level descriptors, largely from the frame-wise estimates described above. It is also extensible to estimate additional descriptors estimated by other packages, either from the frame-wise descriptors or the spectral representation. All of the note-level summary descriptors are added in their own columns in the \verb|nmat| DataFrame, which facilitates both analysis of these parameters within Python and exporting the estimated performance data with note-level linking to the corresponding symbolic data in  MEI format (see Section \ref{sec:exportingData} for details). 

\textbf{Pitch-Related Descriptors}: \texttt{pyAMPACT} calculates five pitch-related descriptors: mean \fo{}, perceived pitch, jitter, vibrato rate, and vibrato depth. Mean \fo{} is calculated as a geometric mean, following from the results reported in \cite{shonle1980pitch}, which indicate that of the simple mean types, the geometric is the closest to the perceived pitch of vibrato tones. A more complex perceived pitch model is also calculated, based on the results reported \cite{gockel2001influence}. This is a weighted mean based on the rate of change, where frames with a slower rate of change in the \fo{} estimates are given more weight. Jitter is approximated by calculating the difference between sequential frame-wise 
\fo{} estimates. Vibrato descriptors are calculated by first computing the spectrum of the note-segmented \fo{} trace with an FFT. Vibrato extent is estimated by doubling the maximum absolute value of the \fo{} trace spectrum and vibrato rate is estimated by finding the position of the maximum absolute value in the spectrum. We are currently working on implementing more sophisticated models of the evolution of pitch and pitch-related parameters over the duration of each note.

\textbf{Dynamics-Related Descriptors}: Mean power is calculated from the note-level frame-wise power estimates using the arithmetic mean. Additionally, shimmer is approximated in an analogous manner to jitter, but by calculating the difference between sequential frame-wise power estimates. Work is currently ongoing to integrate loudness estimation.

\textbf{Timbre-Related Descriptors}: Timbre descriptors are estimated from the harmonic spectral representation derived from the note-wise frame-wise \fo{} and power estimates. Currently, \texttt{pyAMPACT} calculates all of the spectral features available in librosa (spectral bandwidth, spectral centroid, spectral contrast, spectral flatness, and spectral rolloff) and calculates the arithmetic mean of each of these to generate a note-level summary descriptor. Since the spectral representation only models harmonic partials, the accuracy of these descriptors, particularly spectral flatness, is limited. Work is currently ongoing to incorporate the features in the \texttt{Timbre Toolbox} \cite{peeters2011timbre}, including ADSR, harmonic energy, noisiness, inharmonicity, spectral spread, spectral skewness, spectral kurtosis, and spectral crest.

\section{Exporting Data}\label{sec:exportingData}

\texttt{pyAMPACT} includes functions for exporting musical scores with note-linked performance data to MEI format. MEI \cite{roland02} was primarily designed to encode symbolic musical data. The recent inclusion of the \texttt{<extData>} \footnote{\url{https://music-encoding.org/guidelines/v5/elements/extData.html}}  element in the latest release of MEI\footnote{\url{https://music-encoding.org/guidelines/v5/content/introduction.html\#modelChanges}} facilitates the linking of symbolic events (such as notes) to data related to specific time points in linked audio files. \texttt{<extData>} can contain a standard XML \texttt{<![CDATA[]]>} tag, thus the exact specifications of the linked data are flexible. 
Each \verb|<extData>| element wrapped in a \verb|<when>| element, which requires linking to both a specific time-point in an external representation (in our case an audio file) and linking to a specific symbolic event defined in the <score> section of the MEI file (linked by XMLID). \texttt{pyAMPACT} encodes a JSON-formatted object \cite{crockford06} with symbolic data into an \verb|<extData>| element linked to each note in the symbolic representation. 

An example of the MEI encoding is shown in Figure \ref{fig:extData}. Basic MEI data for a single note is shown in the \verb|<score>| element, which includes the XML ID ('note 1'), which is used to link the performance data in the \verb|<when>| element. The onset time of the note in the audio file is specified within the \verb|<when>| stage (absolute="00:00:10:087"). The pitch-, dynamics-, and timbre-related frame-wise and note-level descriptors are encoded within the \verb|<extData>| element. %A more detailed set of example encodings for pop music vocals is available on the pyAMPACT GitHub \footnote{\url{https://github.com/pyampact/examples/}}
% \cite{devaney2023encoding}

%Since \texttt{<extData>} must be linked to a symbolic event, some type of transcription of the audio must be available. However, this could range from a fully specified musical score to note data without rhythmic or metrical information. The approach is flexible, as more detailed score information can be added later if it becomes available. 

%The performance data encoding uses the \verb|<extData>| element \footnote{https://music-encoding.org/guidelines/v5/elements/extData.html}, which allows non-MEI data to be encoded in a \verb|<![CDATA[]]>| tag. 

\section{Availability and Documentation}\label{sec:background}

{pyAMPACT} is available in our GitHub repository\footnote{\url{https://github.com/pyampact/pyampact}} as well as as a pip package\footnote{\url{https://pypi.org/project/pyampact/}}. Function-level Sphinx documentation of \texttt{pyAMPACT} is also available on a GitHub.io site\footnote{\url{https://pyampact.github.io/}}, and a set of Google Colab tutorial notebooks\footnote{\url{https://github.com/pyampact/pyampacttutorials/}} have been developed to help guide users through standard use cases. 
There are currently three Google Colab tutorial notebooks: (1) an introductory notebook that provides an overview of \texttt{pyAMPACT}'s main workflow (corresponding to Section \ref{workflow}); (2) a notebook details how \texttt{pyAMPACT} imports and represents symbolic data (corresponding to Section \ref{sec:symbolic}), (3) a notebook related to multi-modal-processing, which details how annotations are imported and represented in \texttt{pyAMPACT} (corresponding to Section \ref{sec:multimodal}). These tutorials guide users through performing specific tasks with \texttt{pyAMPACT} and provide reusable code that can be integrated into users' own projects. One of our goals as we further develop these tutorials is to make \texttt{pyAMPACT} accessible to musicologists, similar to the way that the \texttt{Humdrum} User Guide\footnote{\url{https://www.humdrum.org/guide/}} and \texttt{HumdrumR}\cite{condit2019humdrumr} vignettes\footnote{\url{https://humdrumr.ccml.gtcmt.gatech.edu/}} do.   

%, (4) a notebook detailing the performance parameter estimation process (corresponding to Section \ref{performanceParamters}), and (5) a notebook detailing how to export musical score data linked with performance data (corresponding to Section \ref{sec:exportingData}.

\section{Conclusions and Future Work}\label{sec:exportingData}

\texttt{pyAMPACT} uses score-audio alignment to link symbolic and audio music representations in order to estimate note-wise frame-level and note-level tuning-, dynamics-, and timbre-related performance descriptors. \texttt{pyAMPACT} can read a range of symbolic formats and can output note-linked audio descriptors/performance data into MEI-formatted files. \texttt{pyAMPACT} also facilitates multi-modal investigations through its infrastructure for linking symbolic representations and annotations to audio (as described in \cite{devaney20using}).

As mentioned above, we are currently working to integrate \texttt{pyAMPACT} with a loudness estimation package to calculate note-wise loudness estimates and to expand the number of timbral descriptors calculated. To facilitate this, we plan to implement a more sophisticated version of the note-wise spectral representation that leverages phase information in order to capture inharmonic partials. We also plan to implement the DTW-hidden Markov model (HMM)-based model included in \texttt{AMPACT} \cite{devaney14estimating}, which estimates individual note onsets and offsets in notated simultaneities. We are also working to support importing under-specified scores, such as annotations from Tony\cite{mauch2015computer}. With under-specified scores, only note onset, duration, and nominal pitch estimates would be required to guide \texttt{pyAMPACT}'s audio processing algorithms, which would expand the repertoire that can be analyzed to music without notated scores. We also plan to support the export of linked performance data to more data formats (such as Humdrum) and explore how \texttt{pyAMPACT} can be more directly integrated with other symbolic music data frameworks (such as DIMCAT\cite{hentschel_introducing_2023} and musif \cite{llorens_musif_2023}). And, finally, since \texttt{pyAMPACT}'s protocol for importing annotations is extensible and can accommodate a range of note-aligned data, we plan to extend this to offer support for motion capture data and response data from psychological studies.  

\section{Acknowledgments}
This work is supported by the National Endowment for the Humanities (NEH) award HAA-281007-21 and National Science Foundation (NSF) award 2228910. The opinions expressed in this work are solely those of the authors and do not necessarily reflect the views of the NEH or NSF.

%%%%%%%%%%%%%%%%%%%%%%%%%%%%%%%%%%%%%%%%%%%%%%%%%%%%%%%%%%%%%%%%%%%%%%%%%%%%%
%bibliography here
\bibliography{icmc2025_paper_template}

\end{document}